\documentclass[12pt]{article}
\usepackage{amsmath,amsfonts,amssymb}
\usepackage{graphics}

\usepackage[latin1]{inputenc}
\usepackage[pdftex]{graphicx}

\usepackage{fullpage}
\usepackage{color}
\usepackage[american]{babel}

\makeatletter \@addtoreset{equation}{section}

\makeatletter\renewcommand\section{\@startsection {section}{1}{\z@}%
                                   {-3.5ex \@plus -1ex \@minus -.2ex}%nn
                                   {2.3ex \@plus.2ex}%
                                   {\normalfont\large\bfseries}}
\renewcommand\subsection{\@startsection{subsection}{2}{\z@}%
                                     {-3.25ex\@plus -1ex \@minus -.2ex}%
                                     {1.5ex \@plus .2ex}%
                                     {\normalfont\bfseries}}

\renewcommand{\baselinestretch}{1.2}

\newcommand{\be}{\begin{equation}}
\newcommand{\ee}{\end{equation}}
\newcommand{\bea}{\begin{eqnarray}}
\newcommand{\eea}{\end{eqnarray}}
\newcommand{\bmat}{\begin{bmatrix}}
\newcommand{\emat}{\end{bmatrix}}

\newcommand{\bbibitem}[1]{\bibitem{#1}\marginpar{#1}}

\def\Label#1{\label{#1}%
  \smash{\hbox to0pt{\raise1ex\hbox{\tiny[#1]}\hss}}}
\def\noLabels{\let\Label=\label}
\def\nobbibitem{\let\bbibitem=\bibitem}

\def\CC{{\cal C}}

\def\CM{{\cal M}}

\def\CO{{\cal O}}

\def\CW{{\cal W}}

\newcommand{\SL}{\mathrm{SL}}

\newcommand{\RR}{\mathbb{R}}
\newcommand{\ZZ}{\mathbb{Z}}

\begin{document}
%\noLabels % uncomment for final production
%\nobbibitem % uncomment for final production

\begin{titlepage}

%\flushright{UCB-PTH-07/nn}
\vfil\

\begin{center}

{\Large{\bf Unitarity Bounds in AdS$_3$ Higher Spin
Gravity}}

\vspace{3mm}

 Alejandra Castro\footnote{email: acastro@physics.mcgill.ca}$^{a}$,
Eliot Hijano\footnote{email: eliot.hijano@mail.mcgill.ca}$^{a}$,
 Arnaud Lepage-Jutier\footnote{email: arnaud.lepage-jutier@mail.mcgill.ca}$^{a}$
\\

\vspace{8mm}

\bigskip\medskip
\smallskip\centerline{$^a$ \it
McGill Physics Department, 3600 rue University, Montr\'{e}al, QC H3A
2T8, Canada}
\medskip
\vfil

\end{center}
\setcounter{footnote}{0}
%%%%%%%%%%%%%%%%%%%%%%%%%%%%%%%%%%%%%%%%%%%%%%%%%%%%%%%%%%%%%%%%%%%%%%%%%%%%%%%%%%%%%%%
\begin{abstract}
\noindent We study $SL(N,\RR)$ Chern-Simons gauge theories in three dimensions. The choice of the embedding of  $SL(2,\RR)$ in $SL(N,\RR)$, together with  asymptotic boundary conditions, defines a theory of higher spin gravity.  Each inequivalent embedding leads to a different asymptotic symmetry group, which we map to an OPE structure at the boundary. A simple inspection of these algebras indicates that  only the $\CW_N$ algebra constructed using the principal embedding could admit a unitary representation for large values of the central charge.

\end{abstract}
%%%%%%%%%%%%%%%%%%%%%%%%%%%%%%%%%%%%%%%%%%%%%%%%%%%%%%%%%%%%%%%%%%%%%%%%%%%%%%%%%%%%%%%%%
\vspace{0.5in}

\end{titlepage}
\renewcommand{\baselinestretch}{1.1}  %Line spacing
%\renewcommand{\arraystretch}{1.5}
%%%%%%%%%%%%%%%%%%%%%%%%%%%%%%%%%%%%%%%%%%%%%%%%%%%%%%%%%%%%%%%%%%%%%%%%%%%%%%%%
%%%%%%%%%%%%%%%%%%%%%%%%%%%%%%%%%%%%%%%%%%%%%%%%%%%%%%%%%%%%%%%%%%%%%%%%%%%%%%%%%%%%%%%%%%%
%\tableofcontents

\newpage
\tableofcontents

\section{Introduction}

Higher spin theories provide a new venue to examine our expectations about quantum gravity.  The pioneer work of Vasiliev gives a background independent formulation of a classical theory of AdS gravity  coupled to an infinite tower of higher spin fields (see e.g. \cite{Vasiliev:2001ur,Bekaert:2005vh} and references within). An immediate consequence is that the gauge symmetries of the theory encompass both diffeomorphisms and the higher spin  transformations, providing a non-linear and non-local theory. This is one of many features that have the potential to address puzzles such as singularity resolution and the significance of black hole horizons. 

In relation to the holographic principle, higher spin theories allow us to investigate in more  depth the dictionary and consequences of this correspondence. For AdS$_4$/CFT$_3$, the first version of the duality was conjectured by Klebanov and Polyakov in \cite{Klebanov:2002ja}, and further refined and tested in e.g. \cite{Sezgin:2002rt, Giombi:2009wh,Giombi:2010vg}. Without going into details, these complicated bulk Vasiliev theories are conjectured to be dual to simple, and  in principle, solvable theories.  This opens the possibility of tracking  the emergence of space-time from the boundary theory, among other effects.  

Our focus here will be in the  three-dimensional version of  AdS higher spin gravity, and hence its two-dimensional dual CFT. This is arguably the simplest setup of the correspondence from the bulk point of view, which has allowed a better understanding of physical phenomena in Vasiliev theory.  Starting with the construction of a classical phase space \cite{Henneaux:2010xg,Campoleoni:2010zq}, the advances include understanding the quantization of the theory \cite{Gaberdiel:2010ar, Castro:2010ce,Bagchi:2011td,Castro:2011ui,Datta:2011za}, a non-geometric definition of black holes \cite{Gutperle:2011kf,Castro:2011fm,Tan:2011tj}, construction of novel solutions \cite{Didenko:2006zd,Ammon:2011nk, Castro:2011iw}, generalizations to de Sitter space \cite{Ouyang:2011fs}, and much more \cite{Chen:2011vp,Chen:2011yx,Campoleoni:2011tn, Gary:2012ms}. Further,  the duality proposed in \cite{Gaberdiel:2010pz} between a specific Vasiliev theory  and a large $N$ 't Hooft limit of  $\CW_N$ minimal models is providing new insights in the field. %\cite{Kiritsis:2010xc,Gaberdiel:2011wb,Ahn:2011pv,Gaberdiel:2011zw,Chang:2011mz,Gaberdiel:2011nt,Campoleoni:2011hg,Bagchi:2011vr,Kraus:2011ds,Papadodimas:2011pf,Ahn:2011by,Ouyang:2011fs,Creutzig:2011fe,Ammon:2011ua,Gaberdiel:2011aa,Chang:2011vk}.
 
The advantage of AdS$_3$ gravity, and its higher spin generalizations, is due to the absence of local degrees of freedom.  The construction of these theories is straightforward by using the Chern-Simons (CS) formulation of 3D gravity, and as we will review below, coupling higher spin fields to gravity is as simple as studying a $SL(N,\RR)\times SL(N,\RR)$ CS theory. While it seems almost trivial, the theory still contains both perturbative and non-perturbative configurations which characterize the global dynamics of the theory.  Our aim is to understand the perturbative spectrum of the higher spin theory, and from here identify which of these classical theories are well-defined after canonical quantization. 

There is a systematic and complete construction of the perturbative spectrum, and the essence of this construction is based on the original work of Brown and Henneaux \cite{Brown:1986nw}. The idea is to construct the non-trivial gauge transformations, and the group generated by this set labels all physical states smoothly connected to the identity. In the bulk language, this is known as the asymptotic symmetry group. Based on \cite{Banados:1994tn,Banados:1998gg}, the analysis in \cite{Campoleoni:2010zq} provides a systematic implementation of the Brown-Henneaux construction adapted to higher spin theories. The remarkable observation is that the resulting algebra is a conformal extension of the Virasoro algebra, known  as $\CW$-algebras \cite{Bouwknegt:1992wg}, with central charge $c$. As in the Drinfeld-Sokolov  reduction \cite{Drinfeld:1984qv} --which is an algebraic  construction of $\CW$-- the resulting  algebra constructed in the bulk depends on the gauge group of the CS theory, the embedding of $sl_2$ in the gauge group and the coupling constant $k$, where $c\sim 6k$. The majority of the literature listed above focus on the principal embedding of $sl_2$ in $sl_N$;  here the gravitational theory has a simple interpretation as an interacting theory for a non-degenerate tower of massless spin $s$ fields with $s=2,\ldots ,N$. Our focus is on the physical interpretation of secondary (non-principal) embeddings, which we will infer by exploiting some basic features of the conformal algebras.

Specific examples of non-principal embeddings have been discussed in the context of higher spin theories in \cite{Ammon:2011nk,Castro:2011fm,Campoleoni:2011hg,Tan:2011tj}. A point that has been overlooked is that the $\CW$-algebra, obtained after imposing asymptotically AdS boundary conditions, is universally ill-defined in the following sense. All non-principal embeddings contain either Abelian or non-Abelian subalgebras generated by spin 1 fields. These subalgebras are enhanced to chiral  Kac-Moody algebras at level $\kappa$ inside the $\CW$-algebra, and $\kappa$ is mostly fixed by the central charge $c$.\footnote{$\kappa$ also depends on the rank of the gauge group and other minor details of the embedding. This has been worked out explicitly in \cite{deBoer:1993iz}.}  Our construction shows that $\kappa$ is strictly negative for large positive values of $c$. As we show explicitly in the text, a negative level implies that the spectrum contains negative norm states. The snapshot of the argument is that the Kac-Moody subalgebra is schematically of the form
\be\label{uuu}
\left[ U_{n},U_{m}\right]
=-|\kappa|n\delta
_{n+m} +\cdots, \ee
and therefore the state  $|\psi\rangle=U_{-1}|0\rangle$ has negative norm. The dots in  \eqref{uuu} are additional terms  appearing for non-Abelian currents, but those terms don't interfere with the logic. The details are discussed  sections \ref{subsec:U1} and \ref{subsec:slN}. 
 Hence, the semiclassical $\CW$-algebra for any secondary embedding does not admit a unitary representation.\footnote{The algebra does admit irreducible representations, unfortunately for large values of $c$ they will be non-unitary.}  However, a class of $SL(N)$ theories that escapes our fatal conclusion are those built using the principal embedding.  Our analysis  provides a simple selection principle that places the principal embedding as perhaps the only consistent framework of higher spin gravity in three dimensions.

%In this work we motivate the selection of the principal embedding of
%$\SL(2,\RR)$ in $\SL(N,\RR)$ as the only working framework of higher
%spin gravity. For this embedding the matter content reproduces the
%aforementioned tower of fields with spin from $2$ to $N$. In any
%other embedding we show that the field content will include a spin $1$
%current that will spoil the unitarity of the theory in the
%semiclassical regime.

The organization of the paper is as follows.  We first work out the matter content of our theory in
section \ref{sec:HS}, focusing on two classes of non-principal
embeddings: the sum and product embedding. We then construct the asymptotic symmetry algebra in
section \ref{sec:OPE} which we use to map our fields to operators at
the boundary. The relevant operator product expansions are then
related to norms of descendents of the spin $1$ current in the
different subsections. In appendix \ref{app:conv} we introduce notation
and conventions, and in appendix \ref{app:QuantumW} we discuss in more depth the unitary representation of $\CW^{(2)}_3$ for finite values of the central charge.

\section{Matter Content of Higher Spin Gravity}\label{sec:HS}

Three-dimensional Einstein gravity with a negative cosmological
constant can be recast as a $\SL(2,\RR)\times \SL(2,\RR)$
Chern-Simons theory \cite{Achucarro:1987vz,Witten:1988hc}.  To fix
some of the notation, we will briefly review this statement here. We
will then proceed to discuss the construction of higher spin
theories using the language of the Chern-Simons theory. See also \cite{Blencowe:1988gj,Coussaert:1995zp,Banados:1998pi,Henneaux:1999ib,Campoleoni:2010zq} for more details  and generalizations of this construction.

The key observation is that by rewriting the dreibein and spin
connection as
\be e=\frac{\ell}{2}(A-\bar{A})\, , \quad
\omega=\frac{1}{2}(A+\bar{A})\, , \ee
with $\ell$ the AdS radius, the  Einstein-Hilbert action can be
written as a $\SL(2,\RR)\times \SL(2,\RR)$ Chern-Simons theory \be
S=S_{CS}[A]-S_{CS}[\bar{A}]\, , \ee where \be \label{CSAction}
S_{CS}[A]=\frac{k}{4\pi}{\rm  tr}\int_{\CM}\left(A\wedge
dA+\frac{2}{3}A\wedge A\wedge A\right)\, . \ee The trace is with
respect to the invariant quadratic form of $\SL(2,\RR)$, and the
integration is over the 3-manifold $\CM$. The level of the Chern-Simons action $k$ is related to the AdS radius ($\ell$) and Newton's constant ($G_3$) by matching the normalization to agree with the Einstein-Hilbert action
\be
\label{kCStolAdS}
k  \,{\rm tr}\left(L_{0}^2\right)=\frac{\ell}{8 G_{3}} \, ,
\ee
where $L_{0}$ is the generator for the center of $\SL(2,\RR)$.

The Einstein's equations are then flatness conditions for the
connections, \be dA +A\wedge A=0\, ,\quad
d\bar{A}+\bar{A}\wedge\bar{A}=0 \, . \ee
%To connect with the metric field, we consider the following combination of the connections:
%\be
%e=\frac{\ell}{2}(A-\bar{A})\, , \quad \omega=\frac{1}{2}(A+\bar{A})\, ,
%\ee
%which are identified as the vielbein and the spin connection respectively, whereas the constant $\ell$ is the AdS radius. The flatness conditions can then be mapped to Einstein equations.
%
The absence of local degrees of freedom for three-dimensional
gravity is evident  in the Chern-Simons formulation.

Coupling matter to this theory can be easily done in this framework.
In particular we can include spin fields by looking at
extensions of the Chern-Simons gauge group.
We require that $\SL(2,\RR)$ sits as a subgroup in this
extension to guarantee gravitational dynamics. Much effort has been
put recently into the study of $\SL(N,\RR)\times \SL(N,\RR)$
 Chern-Simons theory, %\cite{Henneaux:2010xg,Campoleoni:2010zq,Gutperle:2011kf,Ammon:2011nk,Castro:2011fm,Castro:2011ui,Castro:2010ce,Castro:2011iw,Ouyang:2011fs,Tan:2011tj,Campoleoni:2011tn},
 which has been coined higher spin
gravity for reasons that will become clear in our treatment.
The interpretation of this theory in terms of metric-like fields depends
on the choice of embedding of $\SL(2,\RR)$ in $\SL(N,\RR)$, and we will
review here what is known about such embeddings.

For example, consider the case $N=3$.  Here we have only two possible embeddings. The principal embedding of $\SL(2,\RR)$ in $\SL(3,\RR)$ contains a spin $2$ field and spin $3$ field, hence it is a description of Einstein gravity coupled to a spin $3$ field.  The other inequivalent embedding, denoted non-principal embedding, contains
a spin $2$ field -- analogue to the one found in the principal
embedding --  a spin $1$ current, and two bosonic spin $3/2$ fields \cite{Ammon:2011nk}. Even though both theories are loosely speaking ``$\SL(3)$ gravitational theory'', the two inequivalent  embeddings have different matter content, and therefore a different interpretation when written locally in terms of metric-like fields.

The number of inequivalent embeddings and the field content complexity
of the theory increases with $N$, and hence the gravitational interpretation of an $\SL(N)$ Chern-Simons theory is not unique. The number of possible embeddings of the algebra $sl_2$  in $sl_N$
is given by the partition of $N$ \cite{Dynkin:1957um}. We can denote such an
embedding by the branching of the fundamental representation under
the choice of $sl_2$
\be
\underline{N}_{N}\rightarrow\underset{\{j\}}{\oplus}\, n_j \cdot
\underline{2j+1}\, .\ee
Here we have partitioned $N=\sum_{j}n_{j}(2j+1)$; also we denoted the $d$-dimensional
representation of $sl_m$ by $\underline{d}_{m}$ and dropped the
index for $sl_2$.
The centralizer $\CC$ of the embedded $sl_2$ subalgebra will also
play a role, and in this case is given by
 \be
\CC =\underset{\{j\}}{\oplus}\, sl_{n_{j}}\, ,\ee up to $U(1)$
factors.

To illustrate our discussion, let's define the
principal embedding. This is the embedding relevant for the duality proposed by \cite{Gaberdiel:2010ar}, and studied initially in
\cite{Henneaux:2010xg,Campoleoni:2010zq}. We have
\be \underline{N}_{N}\rightarrow \underline{N}\,.
\ee
The centralizer $\CC$ is trivial, and the adjoint
representation $(\underline{ad}_{N}+\underline{1}_{N}\equiv
\underline{N}_{N}\times\underline{\bar{N}}_{N})$ branches as
 \be
\underline{ad}_{N}\rightarrow\underline{3}\oplus\cdots\oplus\underline{2N-1}\,.
\ee
From this branching, one can see the degrees of freedom
organize into a tower of massless fields of spin from $2$ to $N$ \cite{Campoleoni:2010zq}. Therefore one can argue that this embedding gives a description of gravity coupled to a finite tower of  massless higher spin fields.

The non-principal embeddings are more involved since the centralizer
of $sl_2$ into $sl_N$ will be non-trivial. Under $\CC\otimes sl_2$
the adjoint representation will branch as
\be\label{adjoint} \underline{ad}_{N}+\underline{1}_{N}\rightarrow \left[
\begin{array}{c}
\underset{\{i\neq 0\}}{\oplus }\left(
\underline{ad}_{n_{i}}+\underline{1}_{n_{i}}\right) \otimes
\underline{1}+\left(
\underline{n_{0}}\otimes \underline{\bar{n}_{0}}\right) \otimes \underline{1} \\
\underset{\{i\neq 0\}}{\oplus }\underline{1}_{\CC}\otimes \left( \underline{3}%
\oplus ...\oplus \underline{4i+1}\right)  \\
\underset{\{i\neq 0\}}{\oplus }\underline{ad}_{n_{i}}\otimes \left(
\underline{3}\oplus ...\oplus \underline{4i+1}\right)  \\
\underset{\{i\neq j\}}{\oplus }\left( \underline{n_{i}}_{n_{i}}\otimes \underline{\overline{n_{j}}}_{n_{j}}%
\right) \otimes \left( \underline{2\left\vert i-j\right\vert
+1}\oplus ...\oplus \underline{2\left\vert i+j\right\vert +1}\right)
\end{array}%
\right] \, .\ee
% As we will see in the next section, the centralizer leads to a Kac-Moody (KM) subalgebra of the asymptotic symmetry algebra.
%Let us go through the matter content of the theory by
%looking at equation (\ref{adjoint}).
The matter content is given by the branching of $\underline{ad}_{N}$, and thus one of the singlet on the right-hand side of (\ref{adjoint}) will be canceled by the singlet constraint on the left-hand side. The first line represents the
possible spin $1$ fields. The very first term will lead to
non-Abelian currents associated to $\CC$, and $U(1)$ singlets will also be present if $n_{i\neq0}>1$.
The last term in this line will contain a singlet unless $n_0=0$.
The second line in this expression represents singlets under $\CC$ of spin $2$ up to $(2i+1)$, which we identify as the metric
and the higher spin fields in the bulk. The third line represents
multiplets of spin from $2$ to $(2i+1)$ that transform in the
adjoint representation of a $sl_{n_i}$ algebra. The fourth term contains
fields of spin from $(|i-j|+1)$ to $(|i+j|+1)$ that transform
non-trivially under
$\left(\underline{n_{i}}_{n_{i}}\otimes\underline{\overline{n_{j}}}_{n_{j}}\right)$.

We point out that the only embedding that lacks spin $1$ currents
is the one with $n_0=0$ and a unique $n_{i\neq0}\neq0$, which we
single out as the principal embedding. Any other embedding will
contain either singlets of spin $1$ with $U(1)$ gauge symmetry or
multiplets of spin $1$ transforming in the adjoint representation of
a $sl_{n_i}$ algebra.
%This observation is of great importance, as we will show in the following sections that these currents  give rise to negative norm states in the spectrum of the boundary theory.
%We will then conclude that the only embedding that is free of non-unitary states is the principal embedding.

\subsection{More about non-principal embeddings}

The presence of spin $1$ currents will be crucial to understand aspects of these higher spin theories. For sake of simplicity, we will carry out explicit computations for only two class of embeddings: the sum and the product embedding \cite{Bais:1990bs}. The sum embedding will serve as an example for the embeddings with $U(1)$ currents, while the product embedding is a nice setting to study non-Abelian currents.

Taking $N=P+M$, the structure of the sum embedding is $\underline{
P+M}_{P+M}\rightarrow 1\cdot\underline{M}+P\cdot\underline{1}$. Its adjoint decomposition is
\be\label{sumembedding}
\underline{ad}_{P+M}\rightarrow\underline{ad}_P \otimes
\underline{1} + \underline{P}_P \otimes \underline{M} +
 \underline{\overline{P}}_P\otimes \underline{M} + \underline{1}_{P}\otimes\left(\underline{1}\oplus\underline{3}\oplus
\cdots \oplus\underline{2M-1}\right) \, .\ee
The theory contains a $sl_{P}$ algebra, $P$  fields of spin
$(\tfrac{M+1}{2})$  that transform in the fundamental representation of
the $sl_P$ algebra, and another $P$ that
transform in the conjugate representation. Finally we have spin from $1$ to
$M$ fields that transform as singlets under $sl_P$. Note that the latter with spin greater than $1$ can also
be found in the principal embedding of $sl_2$ in $sl_M$
\be
\underline{ad}_{M}\rightarrow
\underline{3}\oplus \cdots
\oplus\underline{2M-1} \, .\ee
%This embedding can be proven to have $W_M$ asymptotic symmetry algebra \cite{}. As a
%consequence, the sum embedding will contain a $sl_P$ KM current
%algebra, a $U(1)$ singlet, and a $W_M$ algebra commuting with the KM
%subalgebra.

 The product embedding will be used to study non-Abelian current
algebras. Here we take $N=PM$, and  the structure is $\underline{
P\cdot M}_{P\cdot M}\rightarrow P\cdot\underline{M}$. Its adjoint
decomposition is
\be\label{productembedding}
\underline{ad}_{P\cdot M}\rightarrow\underline{ad}_P \otimes \underline{1}
+ \underline{1}_{P} \otimes \left(\underline{3}\oplus \cdots \oplus
\underline{2M-1} \right) +
\underline{ad}_{P}\otimes\left(\underline{3}\oplus \cdots \oplus
\underline{2M-1} \right) \, .\ee
The theory contains a $sl_{P}$ current algebra, singlets of spin
from $2$ up to $M$, and a multiplet of $P^2-1$ fields of spins
from $2$ up to $M$ that transform in the adjoint
representation of $sl_P$.

\section{Operator Product Expansions and Unitarity}\label{sec:OPE}

In this section we will describe some general features of the  asymptotic symmetry algebra
for any embedding of $sl_2$ in $sl_N$ through the Drinfeld-Sokolov procedure. Further reference on this topic are \cite{Bowcock:1991zk,Tjin:1993dk,deBoer:1993iz}.   Our goal is to show that the presence of
spin $1$ currents in any embedding implies that the  algebra does not admit a unitary representation in the classical limit.\footnote{By classical limit we mean one for which the central charge of the boundary theory is large, i.e. the AdS radius is large in Planck units. The nature of $\CW$-algebras in a quantum regime has been addressed previously in \cite{deBoer:1993iz}. We use the example of the $\CW_{3}^{(2)}$ algebra in appendix \ref{app:QuantumW} to discuss the possibility of unitary representations for small central charge.}

%We will start by gathering the necessary ingredients needed to describe the

\subsection{Asymptotic symmetry group, $\CW$-algebras and OPEs}

%We start by defining the asymptotic symmetry algebra and its connection with $\CW$-algebras.
The asymptotic symmetry group is the set of non-trivial gauge transformations that preserve specific boundary conditions. For each  generator of the group there is a finite conserved charge associated to it, and the perturbative spectrum of the theory is obtained by acting with these charges. With asymptotically AdS boundary conditions   \textit{\`{a} la} Brown-Henneaux \cite{Brown:1986nw},    the familiar example is pure AdS$_{3}$ gravity.  For this theory the rigid $sl_2\times sl_2$ algebra  is enhanced to two copies of the Virasoro algebra.

A similar statement holds for the higher spin theories:  starting from an $sl_N\times sl_N$ theory with asymptotic AdS boundary conditions, the resulting asymptotic symmetry group is two copies of the $\CW$-algebra \cite{Henneaux:2010xg,Campoleoni:2010zq}.\footnote{ See \cite{Gary:2012ms} for a modification of these boundary conditions.}   By construction these $\CW$-algebras contain the Virasoro generators,  and the additional spin currents depend on the embedding of $sl_2$ in $sl_N$. In this subsection, we will review this construction following the conventions in  \cite{Bais:1990bs, Campoleoni:2010zq}.

In the Chern-Simons formulation of the theory, we first need to specify the connections and topologies under consideration. We start by introducing light-cone coordinates $x^{\pm}=t/\ell\pm \theta$, where $t$ stands for the time direction and $\theta$ parametrizes the circle at the boundary. Constant time slices have the topology of a disc parametrized by $\theta$ and the radial coordinate $\rho$.

We will choose a radial gauge for the connections where
\be A_{\rho}=b^{-1}\partial_{\rho}b~,\quad  \bar A_{\rho}=b\,\partial_{\rho}b^{-1} \, .\ee
This is always possible as there exists a local gauge transformation that brings any possible
 connection to this form. Here $b$ is an arbitrary function of $\rho$
valued in the $SL(N,\RR)$ group. The connection one-forms are now
given by
\be \label{ConnRadial}\begin{split} A&=b^{-1}a(x^{+},x^{-})
b+b^{-1}d\,b \, ,  \\
\bar{A}&=b \,\bar{a}(x^{+},x^{-})b^{-1}+b\,d\,b^{-1}\, ,\end{split}\ee
where $a(x^{+},x^{-})$ and $\bar{a}(x^{+},x^{-})$ are flat $sl_N$ valued one-forms. In order to
ensure the flatness of those one-forms, we impose that
\be  a(x^{+},x^{-}) =a(x^{+})dx^{+}\, , \qquad  \bar{a}(x^{+},x^{-}) =\bar{a}(x^{-})dx^{-}\, .\ee
And from now on, we focus on the connection $A$, since the treatment for $\bar{A}$ is parallel.

Following the discussion of section 2, the $sl_N$ algebra splits
into irreducible representations of $sl_2$, i.e.
$\underline{ad}_N\rightarrow\underset{j}{\oplus}n_{j}\,\underline{2j+1}$.
This means that a generic connection can be specified as
\be \label{ConnParam} a(x^+)={\sum_{j}}\sum_{m=-j}^{j} \Phi^{j,m}_{(n_j)}(x^+)
T_{j,m}^{(n_j)} \, ,\ee
where the $T_{j,m}^{(n_j)}$ are generators combined into representations of weight $j$, and the index $n_j$ labels the different representations of weight $j$.\footnote{According to \eqref{adjoint} the different $T_{j,m}^{(n_j)}$ can have non-trivial relations through the centralizer of $sl_2$. Hence the field is specified by its weight $j$ and its transformation with respect to  the centralizer.} For example, we denote the $sl_2$ generators as $(T_{1,-1}^{(1)},T_{1,0}^{(1)},T_{1,1}^{(1)})$.

We can now define asymptotically AdS boundary conditions. The connection which corresponds to the AdS
background is
\be A_{AdS}=e^{\rho}T_{1,1}^{(1)}dx^{+}+T_{1,0}^{(1)}d\rho \, , \ee
from which we read the radial parameter $b=e^{\rho T_{1,0}^{(1)}}$.  Choosing the same radial parameter for all connections, an asymptotically AdS configuration satisfies
\be
\label{BC} (A-A_{AdS})\vert_{\rho\rightarrow\infty}= \CO(1)\, . \ee
In terms of the expansion \eqref{ConnParam}, this effectively cuts down the fields that would have negative conformal weights, i.e.
\be \Phi^{j,m}_{(n_j)}=0 \quad {\rm for}\quad m>0 \, ,\, j>1\, .\ee
One can then use gauge transformations that respect the boundary
conditions to bring the connection to the highest-weight gauge, i.e.
\be
\label{Afix}
 a= T_{1,1}^{(1)}+a_{fix} \, , \quad  a_{fix}= \sum_{j}\frac{1}{c_j}\Phi^{j}_{(n_j)}T_{j,-j}^{(n_j)}\, ,
 \ee
where $\Phi^{j}_{(n_j)}\equiv \Phi^{j,-j}_{(n_j)} $, and we introduced $c_j={\rm tr}(T_{j,-j}^{(n_j)} T_{j,j}^{(n_j)})$ to assure conventional normalization of the fields.

Despite appearances, there is some residual gauge symmetry. More concretely, consider a gauge transformation
 \be\label{GaugeTransf}
\delta_{\Lambda} a =d\Lambda+\left[a,\Lambda\right] \, ,
 \ee
  with
 \be\label{Lambda}
 \Lambda =\sum_{j}\sum_{m=-j}^{j} \mu_{j,m}^{(n_j)} T_{j,m}^{(n_j)}~.
 \ee
For  some non-trivial relations among the functions $\mu_{j,m}^{(n_j)}$,  this implements only a transformation on $\Phi^{j}_{(n_j)}$. Hence it will preserve conditions \eqref{BC} and \eqref{Afix}.   And for those transformations that do not vanish near the boundary -- physical symmetries -- we  will have the corresponding conserved charges
\be\label{Charge}
Q(\mu)=\frac{k}{2\pi}\int d\theta\, {\rm tr}\left( a_{fix}\Lambda\right)
=\frac{k}{2\pi} \int d\theta {\sum_{j}}\mu_{j}^{(n_j)}\Phi^{j}_{(n_j)} \, ,
 \ee
where $\mu_{j}^{(n_j)}\equiv \mu_{j,j}^{(n_j)}$ is the source conjugate to $\Phi^{j}_{(n_j)}$. The level of the Chern-Simons action is related again to the AdS radius via \eqref{kCStolAdS},
%\be
%\label{kCStolAdS}
%k=\frac{l}{8 G {\rm tr}\left(L_{0}L_{0}\right)} \, ,
%\ee
where now the trace is the bilinear invariant of $sl(N,\RR)$.

Further the Poisson bracket of these charges generate gauge transformation on the fields $\Phi$, i.e.
\be\label{poissonPhi}
\delta_{\mu}\Phi = \{ \Phi, Q(\mu)\} \, ,
\ee
and most importantly
\be\label{poissonQ}
\delta_{\mu'} Q(\mu) = \{ Q(\mu), Q(\mu')\}= Q([\mu,\mu'])+K(\mu,\mu')\, .
\ee
This is the asymptotic symmetry group. For pure AdS$_3$ gravity it
would lead to the Virasoro algebra, where $K$ is the central term.
For the $sl_N$ higher spin theories $K$ contains in addition
nonlinear terms on $Q$. These are the $\CW$-algebras, and examples
of the explicit construction for $N=3,4$ can be found in e.g.
\cite{Campoleoni:2010zq,Campoleoni:2011hg,Tan:2011tj}. For any of
these algebras,  one can then read off the central charge in
\eqref{poissonQ} and recover the celebrated result of Brown and
Henneaux%\footnote{We will illustrate this result in the semiclassical regime in the next sections and we discuss quantum corrections in appendix \ref{app:QuantumW}. For more work on the latter issue we refer the interested reader to \cite{deBoer:1993iz}.} 
\be\label{centralcharge} c=12 k {\rm tr}(L_{0}^2)=\frac{3
\ell}{2 G_{3}} \, , \ee with $L_{0}\equiv T_{1,0}^{(1)}$.

For our purposes it will be convenient to translate
\eqref{poissonPhi} to a statement concerning OPEs of the fields as
in \cite{Gutperle:2011kf}. At the boundary, we will use Euclidean
coordinates $(x^{+},x^{-})\rightarrow(z,\bar{z})$ and use the
holomorphic nature of the fields in our theory. We then use
Noether's theorem to write
\be\label{dO} \delta \Phi (z)=\underset{z\rightarrow
0}{{\rm Res}}[J(z)\Phi(0)] \, ,\ee
where $J(z)=\sum_{j}\mu_{j}^{(n_j)}\Phi^{j}_{(n_j)}$. By adjusting the sources, one can extract from (\ref{GaugeTransf}) and (\ref{dO}) the different OPEs. From simple contour integrations the latter are mapped to commutation relations. In the following sections we use this procedure to
characterize the asymptotic symmetry algebra corresponding to non-principal embeddings.

This procedure is equivalent to the Drinfeld-Sokolov reduction that
leads to extended conformal algebras. One important
ingredient of this procedure, which we will later use in this
section, is the definition of the stress tensor $L$. In the presence of currents, the stress tensor contains a Sugawara density in addition to the spin $2$ field, and in our notation it reduces to
\footnote{ In a general gauge, the stress
tensor could contain a linear improvement term proportional to the
Cartan element of the singled-out $sl_2$ subalgebra. In the
highest-weight gauge the latter drops \cite{Bais:1990bs}.}
 \be\label{sugawara} L=-\frac{k}{2} {\rm  tr}(a^{2}) \, ,\ee
where the proportionality factor is fixed by \eqref{poissonPhi}.
With this definition, the fields $\Phi^{j}_{(n_j)}$ are primaries of weight $(j+1)$.
%The centralizer of the $sl_2$ subalgebra will then be lifted to a Kac-Moody current.

%Before doing any particular computation, let us review here how to
%relate the variation of the fields $\Phi^{k,-k}$ to their OPEs.

\subsection{Sum embedding}\label{subsec:U1}
 We now proceed to apply this rather abstract discussion of the asymptotic symmetry group to a concrete setup.  Consider the sum embedding as defined in \eqref{sumembedding}. Taking $N=M+P$,  we can write the embedding as the branching of the fundamental representation of $sl_{N}$, i.e.
 $$\underline{P+M}_{P+M}\rightarrow 1\cdot\underline{M}+P\cdot\underline{1}~.$$

Constructing the full asymptotic symmetry group is tedious, and not very illuminating. Instead,  we consider a truncation that involves only the gravitational sector and a $U(1)$ current of this embedding. For the sum embedding, the gauge fixed one-form \eqref{Afix} turns out to be
 \be
 \label{afixSum}
 \begin{split} a_{fix} &=\left(
\begin{array}{cc}
-\frac{1}{c_0}\frac{M}{P}{U}\mathbf{1}_{P\times P} & \mathbf{0}_{M\times P} \\
\mathbf{0}_{P\times M} & \frac{1}{c_1}T\, W^{(2)}_{-1}+\frac{1}{c_0}U%
\mathbf{1}_{M\times M}%
\end{array}%
\right) \, .
\end{split}
\ee
with $c_j$ are the traces defined below \eqref{Afix}. Here $T$ is the spin $2$ field and $U$ is
the $U(1)$ current, and stand for the
fields $\Phi^{1}_{(1)}$ and $\Phi^{0}_{(1)}$ respectively. The stress tensor \eqref{sugawara} is given by
\be L=\frac{c}{12}\frac{1}{c_1}\left(2T+\frac{1}{c_0}{U^2}\right)~. \ee
The explicit expressions for the generators, as introduced in \eqref{ConnParam}, are
\be \begin{split}
T_{1,\{0,\pm 1\}}^{(1)}&=\mathbf{0}_{P\times P}\oplus
W^{(2)}_{\{0,\pm 1\}} \, , \\
T_{0}^{(0)}&=\left(-\frac{M}{P}\mathbf{1}_{P\times P}\right)\oplus
\mathbf{1}_{M\times M}\, , \end{split}
 \ee
where $W^{(2)}_{0,\pm 1}$ are traceless $M \times M$ matrices (see appendix \ref{app:conv}).

 To construct the asymptotic symmetry group, it will be sufficient to start with a gauge parameter of the form
\be\label{lambdase}
 \begin{split}
\Lambda  &=\left(
\begin{array}{cc}
-\frac{M}{P}\mu^{(0)}_{0} \mathbf{1}_{P\times P} & \mathbf{0}_{M\times P} \\
\mathbf{0}_{P\times M} & \mu^{(1)}_{1}
W^{(2)}_{1}+\mu^{(1)}_{1,0}W^{(2)}_{0}+\mu^{(1)}_{1,-1}W^{(2)}_{-1}+\mu^{(0)}_{0} \mathbf{1}_{M\times M}%
\end{array}%
\right)\, .\end{split}\ee
Here $\mu^{(1)}_{1}$ and $\mu^{(0)}_{0}$ are the sources for $T$  and $U$, in accordance with \eqref{Charge}. The components $\mu^{(1)}_{1,0}$ and $\mu^{(1)}_{1,-1}$ are determined by requiring that the gauge transformation preserves  \eqref{BC} and \eqref{Afix}. This gives
\be \label{FixExtraLambda} \begin{split}
\mu^{(1)}_{1,0}&=-\partial \mu_{1}^{(1)} \, ,\\
\mu^{(1)}_{1,-1}& = \frac{1}{2}\partial^{2} \mu_{1}^{(1)} + \frac{T}{c_{1}} \mu_{1}^{(1)} \, ,
\end{split} \ee
where we used the notation $\partial \equiv {\partial_{ x^+}}$ and $\bar \partial \equiv {\partial_{x^-}}$ .

As a result, the non-trivial gauge transformations \eqref{lambdase}-\eqref{FixExtraLambda}  act on the fields as
\be
\label{dLdUSum}
\begin{split}
\delta_{\Lambda} L &=\frac{c}{12}\partial ^{3}\epsilon+\epsilon\partial
L+2L\partial\epsilon+U\partial \eta \, ,\\
\delta_{\Lambda} U &=-\frac{M^{2}}{c} \left(\frac{M}{P}+1\right)\left(M^{2}-1\right)\partial
\eta+U\partial\epsilon+\epsilon\partial U \, .
\end{split}
\ee
In \eqref{dLdUSum} we introduced
 \be
\begin{split}
\epsilon &= \mu^{(1)}_{1}\, ,\\
\eta &=-\frac{c }{6c_0c_1 }\left( \mu^{(1)}_{1} U-c_0 \mu_{0}^{(0)}\right) \, .
\end{split}
\ee
in order to get the canonically normalized transformations.

We can reconstruct  the OPEs from \eqref{dLdUSum}. Using $J(z)=\epsilon L+\eta U$ in \eqref{dO} we obtain
 \be
\begin{split}
L\left( z\right) L\left( 0\right) &\sim \frac{c}{2}%
\frac{1}{z^{4}}+\frac{2L\left( 0\right)}{z^{2}}+\frac{\partial L\left( 0\right)}{z} \, , \\
L\left( z\right) U\left( 0\right) &\sim \frac{U\left( 0\right)}{z
^{2}}+\frac{\partial U\left( 0\right)}{z}  \, ,\\
 U\left(
z\right) U\left( 0\right) &\sim
-\frac{M^2}{c}\left(\frac{M}{P}+1\right)\left(M^2-1\right)\frac{1}{z^{2}}
\, .
\end{split}
 \ee
Decomposing $L$ and $U$ in a Laurent expansion,  the singular parts of the operator expansions yield commutation
relations by contour integration
 \be
 \label{CommuSum}
 \begin{split}
 \left[
L_{n},L_{m}\right]  &=\left( n-m\right) L_{n+m}+\frac{c}{12}\left(
n^{3}-n\right) \delta _{n+m} \, , \\
 \left[ L_{n},U_{m}\right]  &=-mU_{m+n} \, , \\
 \left[ U_{n},U_{m}\right]
&=-\frac{M^2}{c}\left(\frac{M}{P}+1\right)\left(M^2-1\right)n\delta
_{n+m} \, ,
\end{split}
 \ee
where $n,m\in\ZZ$.

Even though this is not the complete asymptotic algebra, we can make some precise assertions about the entire Fock space. It is clear from \eqref{CommuSum} that for $c>0$, which is our case due to \eqref{centralcharge}, the level of the $U(1)$ current is negative. This immediately indicates that the spectrum is sick: by simple inspection negative norm states will be present in the theory. More explicitly, in a highest weight representation,  the vacuum is defined as
\be
L_n|0\rangle =U_n|0\rangle=0 ~,\quad n\ge 0~,
\ee
and analogous expressions for the other spin generators. The above also implies that the vacuum is annihilated by the rigid $sl_N$ generators, and in particular $L_{-1}|0\rangle=0$. Descendants are constructed by acting on the vacuum with  creation operators associated to each spin generator, e.g. $L_{-n-1}$ and $U_{-n}$ with $n>0$.

The first excited state is given solely by
$|\psi\rangle=U_{-1}|0\rangle$; note that this is true only for the vacuum state and it is unaffected by the other spin generators with $j\neq0$.\footnote{The presence of additional states at level 1 would require computing the Kac determinant, and from there assure that the matrix is positive definite. In our setup, this will  be the case if and only  if we have additional spin $1$ fields, and it is the subject of  section \ref{subsec:slN}.} According to \eqref{CommuSum}, the state has norm
\be\label{ppinner}
\langle\psi|\psi\rangle=-\frac{M^2}{c}\left(\frac{M}{P}+1\right)\left(M^2-1\right)\langle 0|0\rangle<0~,
\ee
where $(U_n)^\dagger=U_{-n}$. Hence, the physical spectrum associated to this embedding does not admit unitary representations. As a gravitational higher spin theory, the presence of $U(1)$ fields makes the theory pathological. Note that we could have chosen $(U_n)^\dagger=-U_{-n}$, and then the inner product \eqref{ppinner} is positive definite. At this stage this is compatible with reality of the algebra \eqref{CommuSum}, and it is a reflection of invariance under $U_n\to i U_n$. However, the remaining fields with $j\neq 0$ fix the signature of the $UU$ OPE by imposing a reality condition of their $U(1)$ charge. This will be explicit in the example below. The inner product defined here is the only one compatible with reality of the entire $\CW$-algebra.  

This conclusion is generic to any embedding containing $U(1)$ currents, not only the sum embedding. Including other fields with $j\neq 0$ in the branching of the adjoint representation of $SL(N,\RR)$ \eqref{adjoint}  does not affect the $UU$ OPE; this OPE depends only on the variation $\delta U$ with respect to its associated source $\eta$. In the next sub-section, we will generalize the analysis for multiple currents, but first we will work out an explicit example.

\subsubsection{Example: $U(1)$ singlet in $\CW_{3}^{(2)}$ algebra}\label{subsec:W32}
One example of the sum embedding that can be carried out explicitly --without leaving any field behind-- is the diagonal embedding of $SL(2,\RR)$ in $SL(3,\RR)$. The partition is $\underline{3}\rightarrow 1\cdot\underline{2}+1\cdot
\underline{1}$ and the adjoint representation of the algebra is
\be \underline{ad}_3\rightarrow \underline{1}\otimes\underline{2} +
\underline{1}\otimes\underline{2} +
\underline{1}\otimes\underline{3}+ \underline{1}\otimes\underline{1}
\, . \ee
The theory contains two bosonic spin $3/2$ fields, a spin $2$ field that
carries the gravitational dynamics, and a $U(1)$ current. One can do the same analysis as we have done from \eqref{afixSum} to \eqref{CommuSum}, by simply setting $P=1,M=2$.

In the highest weight gauge we can represent the one-form \eqref{Afix}\footnote{In \cite{Ammon:2011nk}, a different representation was used. In their notation we would have \be\label{Afix32} a_{fix, there}=\left(
\begin{array}{ccc}
j & \bar{G} & T \\
0 & -2j & G \\
0 & 0 & j%
\end{array}%
\right) \, .
 \ee}
\be\label{aw32} a_{fix}=\left(
\begin{array}{ccc}
-2j &0 & G \\
\bar{G} & j & T \\
0 & 0 & j%
\end{array}%
\right) \, ,
 \ee
where $j$ is the $U(1)$ current, $G$ and $\bar{G}$ are the spin
$3/2$ fields and $T$ is the graviton. They play the role of
$\Phi^{(1)}_{0}$, $\Phi^{(1)}_{1}$, $\Phi^{(1)}_{\frac{1}{2}}$ and
$\Phi^{(2)}_{\frac{1}{2}}$ in
equation \eqref{Afix}.
%The conjugate sources are defined such that
%
 %\be
%{\rm  tr}\left( a_{fix}\Lambda\right)
%=T\epsilon+j\gamma+G\rho+\bar{G}\bar{\rho}\, .
 %\ee
%In \cite{Ammon:2011nk}, a further rescaling of the fields is performed together with defining the stress tensor \eqref{sugawara}. We list here this rescaling and the shift in the source for the field $j$
%
%\be
%\begin{split}
%T &=-\frac{6}{c}\left[ L+\frac{9}{2c}U^{2}\right] \, ,  \\
%j &=\frac{3}{c}U \, , \\
%\gamma &=-\frac{1}{3}\eta+\frac{3}{c}U\epsilon \, , \\
 %G &=-\frac{6}{c} G_{-} \, , \\
 %\bar{G} &= -\frac{6}{c} G_{+} \, .
%\end{split}\ee
%
The OPE was computed in \cite{Ammon:2011nk} using the techniques outlined above. We will not repeat the derivation here and just state  the final answer found in   \cite{Ammon:2011nk}
 \be\label{w32c}
\begin{split}
L\left( z\right) L\left( 0\right) &\sim \frac{c}{2}%
\frac{1}{z^{4}}+\frac{2L(0)}{z^{2}}+\frac{\partial L(0)}{z} \, , \\
L\left( z\right) U\left( 0\right) &\sim \frac{U(0)}{z
^{2}}+\frac{\partial U(0)}{z} \, , \\
 U\left(z\right) U\left( 0\right) &\sim -\frac{c}{9}\frac{1}{z^{2}} \, , \\
 L\left( z\right) G_{\pm }\left(
0\right) &\sim \frac{3}{2}\frac{G_{\pm}(0)}{z^{2}}+\frac{\partial G_{\pm}(0)}{z
} \, , \\
G_{+}\left( z\right) G_{-}\left( 0\right) &\sim
-\frac{c}{3}\frac{1}{
z^{3}}+\frac{3U(0)}{z^{2}}-\frac{1}{%
z}\left( L(0)+\frac{18}{c}U(0)^{2}-\frac{3}{2}\partial U(0)\right) \, , \\
G_{\pm }\left( z\right) U\left( 0\right) &\sim \mp \frac{G_{\pm}(0)}{z}
\, ,
\end{split}
 \ee
where
\be
\begin{split}
T &=-\frac{6}{c}\left[ L+\frac{9}{2c}U^{2}\right] \, ,  \quad j =\frac{3}{c}U \, , \\
 G &=-\frac{6}{c} G_{-} \, , \quad
 \bar{G} = -\frac{6}{c} G_{+} \, .
\end{split}\ee
This is known as the $\CW^{(2)}_{3}$ or Polyakov-Bershadsky algebra \cite{Polyakov:1989dm,Bershadsky:1990bg}. The relevant structure of $L$ and $U$ is preserved after  the inclusion of the spin $3/2$ fields. In particular notice that the negative sign in the $UU$ OPE cannot be removed by a field redefinition without affecting other properties of the algebra. For instance, taking $U\to i U$ spoils the reality of the $U(1)$ charge carried by $G$. Therefore representations of the $\CW^{(2)}_{3}$ algebra will contain negative norm states, as it was argued above for the general sum embedding.

Note that using the Chern-Simons theory we will always obtain the classical limit (large central charge limit) of the $\CW$-algebra. Some of these algebras have a known quantum version, and the claims about unitary representations we make here could change. We refer the interested reader to appendix \ref{app:QuantumW}, where we compute the  Kac determinant for the first levels of the quantum $\CW^{(2)}_{3}$ algebra.

\subsection{Product embedding}\label{subsec:slN}
 We consider now an embedding of $sl_2$ in $sl_N$ that
contains non-Abelian currents. A good representative is the product embedding, where the partition of
$N=P\cdot M$  is
$\underline{P\cdot M}_{P\cdot M}\rightarrow P\cdot \underline{M}$ and the adjoint
representation of the algebra is explained below
equation \eqref{productembedding}.

For sake of simplicity, we turn off the fields that transforms in $\underline{ad}_{P}\otimes\left(\underline{3}\oplus\cdots\oplus\underline{2M-1}\right)$. The connection will only contain the graviton and non-Abelian currents, i.e. we choose \eqref{Afix} as
\be \label{PExyz} a_{fix}=\frac{1}{c_{0}}U^a\sigma_a
\otimes \mathbf{1}_{M\times M}+\frac{1}{c_1 }T\mathbf{1}_{P\times P}\otimes
W^{(2)}_{-1} \, .\ee
The fields $U^{a}$ are currents  that transform under $sl_P$ and $T$  is again the gravitational spin $2$ field. We pick a representation of the $sl_P$ algebra $\lbrace\sigma _{a}\rbrace$ such that
\be
\frac{{\rm tr}\left(\sigma_{a}\sigma_{b}\right)}{\left(\sum_{c}{\rm tr}\left(\sigma_{c}^{2}\right)\right)}=\gamma_{ab} \, , \quad \left[\sigma_{a},\sigma_{b}\right]={f_{ab}}^{c}\sigma_{c} \, ,
\ee
where $\gamma_{ab}$ is the Killing form and ${f_{ab}}^{c}$ are the structure constants. The explicit expression for the generators are
\be \begin{split}
T_{0}^{(a)}&=\sigma_{a}\otimes \mathbf{1}_{M\times M} \, , \\
T_{1,\{0,\pm 1\}}^{(1)}&=\mathbf{1}_{P\times P}\otimes
W^{(2)}_{\{0,\pm 1\}}\, . \end{split}
\ee
To ease the notation, we modified slightly the normalization of the $sl_{P}$ currents to
\be
c_{0} = \sum_a{\rm tr}(T_{0}^{(a)}T_{0}^{(a)})=M\sum_a{\rm
tr}\left(\sigma^2_a\right)~.
\ee
The stress tensor is then \eqref{sugawara}
 \be
 \label{stressProduct}
 L=\frac{c}{12}\frac{1}{c_1}\left(2T+\frac{1}{c_0}{U^a U^b \gamma_{ab}}\right)\, . \ee

The gauge parameter that preserves our boundary conditions takes the form
\bea\label{LLPE} \Lambda
&=&\mu^{(a)}_{0}\sigma_{a}\otimes \mathbf{1}_{M\times M}+\mu^{(1)}_{1}
\mathbf{1}_{P\times P}\otimes W^{(2)}_{1}\cr &&+
\mu^{(1)}_{1,0}\mathbf{1}_{P\times P}\otimes W^{(2)}_{0,M\times M}+\mu^{(1)}_{1,-1}\mathbf{1}%
_{P\times P}\otimes W^{(2)}_{-1}
 \, , \eea
where $\mu^{(a)}_{0}$ and $\mu^{(1)}_{1}$ are sources, and  $\mu^{(1)}_{1,0}$ and $\mu^{(1)}_{1,-1}$ are again simply given by \eqref{FixExtraLambda}.

 The variation of the fields under residual gauge transformations \eqref{LLPE} can be written as
\be
\begin{split}
\delta_\Lambda L &=\frac{c}{12}\partial ^{3}\epsilon +2L\partial \epsilon
+\epsilon
\partial L+U^{a}\partial \eta ^{b}\gamma _{ab} \, , \\
\delta_\Lambda U^{a} &=-\frac{MP}{c}\left(M^2-1\right)c_0
\partial \eta ^{a}+\partial \left( \epsilon U^{a}\right)
-\frac{MP}{c}\left(M^2-1\right){f^{a}}_{bc}U^{b}\eta^{c} \, .
\end{split} \ee
Again, we redefine the sources as
\be
\label{redefSourcesProduct}
\begin{split}
\epsilon&= \mu^{(1)}_1\, ,\\
\eta^{a}&=-\frac{c}{6c_0 c_1} \left( \mu^{(1)}_1 U^a -c_0 \mu^{a}_{0} \right)\, ,
\end{split}
\ee
in order to get the transformations in a canonical form.

Using $J=\epsilon L + \eta_a U^a$, the above transformations completely determines the singular part of the OPEs
\be
\begin{split}
L\left( z\right) L\left( 0\right) &\sim \frac{c}{2}%
\frac{1}{z^{4}}+\frac{2 L \left( 0\right)}{z^{2}}+\frac{\partial L\left( 0\right)}{z}  \, , \\
L\left( z\right)U^{a}\left( 0\right) &\sim
\frac{U^{a}\left( 0\right)}{z ^{2}}+\frac{\partial
U^{a}\left( 0\right)}{z} \, , \\
U^{a}\left( z\right) U^{b}\left( 0\right) &\sim
-\frac{M^2P}{c}\left(M^2-1\right)\frac{\gamma
^{ab}}{z^{2}}+\frac{MP}{c}\left(M^2-1\right)\frac{{f^{ab}}_{c}U^{c}\left(
0\right)}{z } \, ,
\end{split} \ee and the following commutation relations
 \be
 \begin{split}
\left[
L_{n},L_{m}\right]  &=\left( n-m\right) L_{n+m}+\frac{c}{12}\left(
n^{3}-n\right) \delta _{n+m}  \, , \\
\left[ L_{n},U^{a}_{m}\right]  &=-mU^{a}_{m+n}  \, , \\
 \left[ U^{a}_{n},U^{b}_{m}\right]
&=-\frac{M^2P}{c}\left(M^2-1\right) \, n\delta
_{n+m}\gamma^{ab}+\frac{MP}{c}\left(M^2-1\right)f^{ab}_{~~c}U^{c}_{m+n}
\, .
\end{split} \ee
where it is clear that the  $sl_P$ currents form now a Kac-Moody algebra.

One can check that the Kac-Moody currents are problematic, providing
with a generalization of the single $U(1)$ case.  To construct the
spectrum of the theory, we follow the arguments at the end of
section \ref{subsec:U1}. At level 1, we now have a collection of the
states $|\psi^a\rangle=U^{a}_{-1}|0\rangle$, and the norms of these
states are given by \be
\langle\psi^a|\psi^b\rangle=-\frac{M^2}{c}\left(\frac{M}{P}+1\right)\left(M^2-1\right)\gamma^{ab}\langle
0|0\rangle~. \ee The Killing matrix $\gamma_{ab}$ generically will
contain both positive and negative eigenvalues, hence the matrix is
not positive definite. Just as in the sum embedding case, the hermiticity  is unambiguous when one take into account the OPE of other fields with the spin 1 currents.

This captures the basic pathology of having  spin $1$
fields in  $SL(N)$ higher spin theories. We conclude
that any non-principal embedding will contain negative norm states
for positive central charge $c$. We next present an explicit example
of the simplest product embedding, keeping track of the inclusion of
all the fields that are present in the adjoint representation.

\subsubsection{Example: $sl_{2}$ currents in $\CW_{4}^{(2,2)}$ algebra}\label{subsec:W422}
In this section we check that the truncation for the product embeddings does not interfere with the results for the non-Abelian currents. We consider the decomposition $\underline{4}_{4}\rightarrow 2 \cdot
\underline{2}$. The field content consists of a $sl_{2}$ current, a
multiplet of $3$ spin $2$ fields that transform under the
adjoint representation of $sl_{2}$, and a spin $2$ field in the trivial representation of $sl_{2}$
that we will call the graviton. We denote the asymptotic algebra $\CW_{4}^{(2,2)}$ which is the simplest instance of a non-principal, non-diagonal embedding. Note that this algebra is the case $P=2, M=2$ of a product embedding.

The gauge fixed current can be written as
\be
a_{fix}=\left(
\begin{array}{cc}
\mathbf{J} & \mathbf{T} \\
\mathbf{0} & \mathbf{J}%
\end{array}%
\right) \, , \quad  \mathbf{J}=\frac{U^{a}}{2 \sum_{c}\left({\rm tr}\left(\sigma_{c}^{2}\right)\right)}\sigma_{a} \, , \quad \mathbf{T}=-\frac{T}{4}\mathbf{1}+\frac{t^{a}}{2 \sum_{c}\left({\rm tr}\left(\sigma_{c}^{2}\right)\right)}\sigma_{a} \, .\ee
The fields $T,U^{a}$ and $t^{a}$ are the graviton, the $sl_2$ currents and the multiplet of spin $2$ fields respectively. The gauge parameter is now written as follows
\be
\Lambda =\left(
\begin{array}{cc}
\gamma +\tfrac{1}{2}\lambda_{0} & \lambda_{-1} \\
\lambda_{1}  & \gamma -\tfrac{1}{2}\lambda_{0}%
\end{array}%
\right) \, , \quad \gamma=\mu_{0}^{(a)}\sigma_{a} \, , \quad \lambda_{i}=\mu_{1,i}^{(1)}\mathbf{1}+\mu_{1,i}^{a}\sigma_{a}\, . \ee
The conjugate sources are encoded in $\lambda_1$ and $\mu_{0}^{(a)}$. And  $\lambda_{0,-1}$ are solved for in a manner similar to \eqref{FixExtraLambda}.

The stress tensor \eqref{stressProduct} is not affected by the multiplet of spin $2$ fields $t^a$, nor is the redefinition of the sources \eqref{redefSourcesProduct}. We include here the variations of the stress tensor and the currents under the residual gauge transformations
\be
\label{dLdUW422}
\begin{split}
\delta L &=\frac{c}{12}\partial ^{3}\epsilon +2L\partial \epsilon
+\epsilon
\partial L+U^{a}\partial \eta ^{b}\gamma _{ab} +2 t^{a}\partial\mu_{1}^{b}\gamma_{ab} \, , \\
\delta U^{a} &=-\frac{4}{c}c_0
\partial \eta ^{a}+\partial \left( \epsilon U^{a}\right)
-\frac{4}{c}{f^{a}}_{bc}U^{b}\eta^{c}+{f^{a}}_{bc} t^{b}\mu_{1}^{c} \, .
\end{split}
\ee
In particular, we have that the $t^a$ fields are indeed spin $2$ multiplets transforming in the adjoint of the $sl_2$ currents. Moreover, one can see that the $U^{a}U^{b}$ OPE will not be affected by the multiplets, hence the same type of states as in the general case will preclude unitary representations of the $\CW_{4}^{(2,2)}$ algebra.

\section*{Acknowledgements}

We are very grateful to Alex Maloney and  Max Riegler  for useful discussions and comments. We also thank Jan de Boer, Daniel Grumiller, Michael Gutperle and Per Kraus for discussions. This work was supported by the National Science and Engineering Research Council of Canada. E.H. acknowledges support from Fundaci\'on Caja Madrid.

\appendix

\section{Conventions}\label{app:conv}

An explicit $M$-dimensional representation of the  $sl_M$ generators is \cite{Castro:2011iw}
 \bea
W^{(2)}_{1}&=&-\left(
\begin{array}{cccccc}
0 & 0 & 0 & 0 & 0 & 0 \\
\sqrt{M-1} & 0 & 0 & 0 & 0 & 0 \\
0 & \ddots  & 0 & 0 & 0 & 0 \\
0 & 0 & \sqrt{i\left( M-i\right) } & 0 & 0 & 0 \\
0 & 0 & 0 & \ddots  & 0 & 0 \\
0 & 0 & 0 & 0 & \sqrt{M-1} & 0%
\end{array}\right)\\
W^{(2)}_{-1}&=&\left(
\begin{array}{cccccc}
0 & \sqrt{M-1} & 0 & 0 & 0 & 0 \\
0 & 0 & \ddots & 0 & 0 & 0 \\
0 & 0  & 0 & \sqrt{i\left( M-i\right) } & 0 & 0 \\
0 & 0 & 0 & 0 & \ddots & 0 \\
0 & 0 & 0 & 0  & 0 & \sqrt{M-1} \\
0 & 0 & 0 & 0 & 0 & 0%
\end{array}
\right)\\
W^{(2)}_{0}&=&\frac{1}{2}\left(
\begin{array}{cccccc}
M-1 & 0 & 0 & 0 & 0 & 0 \\
0 & M-3 & 0 & 0 & 0 & 0 \\
0 & 0  & \ddots  & 0 & 0 & 0 \\
0 & 0 & 0 & M+1-2i  & 0 & 0 \\
0 & 0 & 0 & 0  & \ddots & 0 \\
0 & 0 & 0 & 0 & 0 & -(M-1)%
\end{array}
\right) \eea
 and
 \be W_{m}^{(s)}=\left( -1\right)
^{s-m-1}\frac{\left( s+m-1\right) !}{\left(
2s-2\right) !}\underset{s-m-1\text{ terms}}{\underbrace{[W^{(2)}_{-1},[W^{(2)}_{-1},%
\ldots ,[W^{(2)}_{-1}}},(W^{(2)}_{1})^{s-1}]\ldots ]]\, .\ee
with  $m=-(s-1)\ldots (s+1)$ and $s=3\ldots M$.

The Killing metric of this algebra is
\be\label{trace}
{\rm  tr}\left(W_{m}^{(s)}W_{n}^{(r)}\right)=t^{(s)}_m\delta^{r,s}\delta_{m,-n}
\, ,\ee where \be\label{tOfTrace}
t^{(s)}_m=(-1)^m\frac{(s-1)!^2(s+m-1)!(s-m-1)!}{(2s-1)!(2s-2)!}M
\prod^{s-1}_{i=1}\left(M^2-i^2\right) \, .\ee

\subsection{Sum embedding construction}\label{app:SE}
The generalization of \eqref{afixSum} to include the remaining matter fields is
\be
a_{fix}=\left(
\begin{array}{cc}
\mathbf{J}_{P\times P}-\frac{M}{P c_0}U\mathbf{1}_{P\times P} &
\begin{array}{cc}
 \mathbf{0}_{\left( M-1\right) \times P} & \mathbf{G}_{1\times P}%
\end{array}
\\
\begin{array}{c}
\overline{\mathbf{G}}_{P\times 1} \\
\mathbf{0}_{P\times \left( M-1\right) }%
\end{array}
& \mathbf{W}_{M\times M}+\frac{1}{c_0}U\mathbf{1}_{M\times M}%
\end{array}%
\right) \, .\ee Here $\mathbf{J}_{P\times P}$ contains the $sl_{P}$ currents, $\mathbf{G}$ ($\mathbf{\overline{G}}$) is the multiplet of
fields that transform in the fundamental (conjugate) representation
of $sl_P$ and $U$ is the spin $1$ singlet. $\mathbf{W}$ contains the
information for the rest of higher spin fields
\be \mathbf{W}_{M\times
M}=W^{(2)}_{1}+\sum_{s=2}^{M}\frac{\Psi
_{s}}{c_s} W_{-s+1}^{\left( s\right) }\, ,\ee where
$W_{s-1}^{\left(s\right)}$ are generators of $sl_M$, and
\be\label{tracescs}
c_s={\rm  tr}\left(W_{s-1}^{\left( s\right) }W_{-s+1}^{(s) }\right)~,\quad c_0= {\rm  tr}\left(T_{0}^{\left( 0\right) }T_{0}^{(0) }\right)~.
\ee

\subsection{Product embedding construction}\label{app:PE}
The generalization of the gauge fixed one-form \eqref{PExyz}  is
 \be a_{fix}= \mathbf{J}_{P\times
P}\otimes\mathbf{1}_{M\times M} +\mathbf{1}_{P\times P}\otimes
L_{1,M\times M} + \sum_{s=2}^{M}\frac{1}{c_s}\mathbf{\Psi} _{s,P \times
P}\otimes W_{-s+1,M\times M}^{\left( s\right) }
 \, . \ee
 Here $\mathbf{1}$ is the unit matrix, $\mathbf{J}$ is a traceless
 matrix that contains the $sl_{P}$ currents, and $\mathbf{\Psi} _{s}$ contains both the spin $s$ field and the multiplets of $P^2-1$ spin $s$ fields that
 transform in the adjoint representation of the $sl_{P}$ currents. Any
 of these matrices can be decomposed as
 $\mathbf{\Psi}_s=\frac{\Psi_s}{P}\mathbf{1}_{P \times P}+\psi^a_s\sigma_{a}$ where
 $\sigma_{a}$ are generators of the $sl_P$ algebra.

\section{Quantum $\CW^{(2)}_3$  algebra}\label{app:QuantumW}
The algebras we have used in this work are semiclassical limits of
$\CW$-algebras. The full quantum algebra receives corrections that affect the relation between the level of the current algebras and the central charge; it also affects the coefficients of non-linear terms. An illustrative example is
the  Polyakov-Bershadsky algebra
$\CW^{(2)}_3$ \cite{Polyakov:1989dm,Bershadsky:1990bg}. Here we will explore if our conclusions in section \ref{subsec:W32} are modified in the quantum regime.

In the quantum regime, the matter content does not change, but the commutation relations get modified slightly. We have
\be
\begin{split}
\label{commW32}
\left[ L_{n},U_{m}\right]  =&-mU_{m+n} \, , \\
 \left[
L_{n},L_{m}\right]  =&\left( n-m\right) L_{n+m}+\frac{c}{12}\left(
n^{3}-n\right) \delta _{n+m} \, \\
 \left[ U_{n},U_{m}\right]
=&\frac{2\kappa+3}{3}n\delta _{n+m} \, , \\
\left[ L_{n},G_{r}^{\pm }\right] =&\left( \frac{n}{2}-r\right)
G_{n+r}^{\pm
} \, ,\\
 \left[ U_{n},G_{r}^{\pm}\right]
=&\pm G^{\pm}_{r+n} \, , \\
\left[ G_{r}^{+},G_{s}^{-}\right] =&\frac{\left( \kappa+1\right)
\left(
2\kappa+3\right) }{2}\left( r^{2}-\frac{1}{4}\right) \delta _{r+s,0} \\
&+\frac{3}{2}\left( \kappa+1\right) \left( r-s\right) U_{r+s}-\left(
\kappa+3\right) L_{r+s}+3\sum\limits_{m}:U_{r+s-m}U_{m}:\, ,
\end{split}
\ee
where $(n,m)\in \mathbb{Z}$ and $(r,s)\in (\mathbb{Z}+1/2)$;  the notation $: \,:$  denotes annihilation-creation normal ordering. The central charge of the  algebra is corrected by quantum effects
\be
\label{quantumC}
c=25-\frac{24}{\kappa+3}-6\left(\kappa+3\right) =-{(2\kappa+3)(3\kappa+1)\over (\kappa +3)}\, ,
\ee
and $\kappa$ is a real parameter. The classical limit is given by $-\kappa\gg 1$, reproducing the OPEs \eqref{w32c}.

In the highest-weight representation, the vacuum satisfies
\be
L_{n}\vert 0 \rangle=0 \quad , \quad U_{n} \vert 0 \rangle =0 \quad , n\ge 0 ~,
\ee
and
\be
 G_{r}^{\pm}\vert 0 \rangle =0 \,\quad r\ge -\frac{1}{2} .
\ee
From the algebra it is also true that $L_{-1}\vert 0 \rangle=0 $.

We define the hermitian conjugate as 
\be L_{-n}^{\dagger}\equiv
L_{n} \quad ; \quad  U_{-n}^{\dagger}\equiv U_{n} \quad ; \quad
(G_{-n}^{+})^{\dagger}\equiv G_{n}^{-}\, . \ee
At level 1 the only
state is $U_{-1}|0\rangle$ and its norm reads 
\be \langle
0|U_{1}U_{-1}|0\rangle=\left(2\kappa+3\right)/3 \, . \ee 
Demanding non-negative norms, implies that the allowed values are $\kappa\ge-{3\over2}$.

The next level is as simple as the first, since the only states are $G_{-3/2}^{\pm}|0\rangle$. The norm is then 
\be
\langle0|(G^{\pm}_{-3/2})^{\dagger}G^{\pm}_{-3/2}|0\rangle=\mp\left(2\kappa+3\right)(\kappa+1)\,
. \ee 
At this stage we see both states cannot have simultaneously positive norm unless 
\be\label{range}
\kappa=-{3/2}~,\quad {\rm or}\quad \kappa=-1~.
\ee

%Imposing positive norms at the two lowest level, the allowed values for $\kappa$ are
%
% \be \label{range} -\frac{3}{2} < \kappa <
%-\frac{11}{12} \, . \ee 
%
%Note that in this range $c$ is valued
%between $0$ and $1$, and therefore this  bound is not consistent with a semiclassical limit.

We now compute the inner product matrix at level 2, in the basis given by $L_{-2}\vert 0 \rangle$, $U_{-2}\vert 0 \rangle$ and $U_{-1}U_{-1}\vert 0 \rangle$. Using the commutation relations \eqref{commW32} we get
\be
\left(\begin{array}{ccc}
\frac{c}{2} & 0 & \frac{2\kappa+3}{3} \\
0 & 2 \frac{2\kappa+3}{3} & 0 \\
\frac{2\kappa+3}{3} & 0  &2 \left(\frac{2\kappa+3}{3}\right)^{2}
\end{array}\right) \, .
\ee
For the two allowed values of $\kappa$ in \eqref{range}, this matrix has non-negative eigenvalues. In particular for $\kappa=-3/2$ we have $c=0$ and all 3 states at level 2 are null; for  $\kappa=-1$  the central charge is $c=1$ and one state is null. 
We do not intend to compute the Kac determinant for every level or other primary states; we leave for the adventurous reader to explore if the limiting cases $c=0,1$ allow for unitary representations.

\bibliographystyle{utphys}

\bibliography{all}

\end{document}